\documentclass[lettersize,journal]{IEEEtran}
\usepackage{amsmath,amsfonts}
\usepackage{algorithmic}
\usepackage{algorithm}
\usepackage{array}
\usepackage[caption=false,font=normalsize,labelfont=sf,textfont=sf]{subfig}
\usepackage{textcomp}
\usepackage{stfloats}
\usepackage{url}
\usepackage{verbatim}
\usepackage{graphicx}
\usepackage{cite}

\usepackage{hyperref}
\usepackage{multirow}
\usepackage{booktabs}
\usepackage{tabularx}

\begin{document}

\title{RoWSFormer: A Robust Watermarking Framework with Swin Transformer for Enhanced Geometric Attack Resilience}

\author{Weitong Chen, Yuheng Li}

\markboth{Journal of \LaTeX\ Class Files,~Vol.~14, No.~8, August~2021}%
{Shell \MakeLowercase{\textit{et al.}}: WMFormer}

\IEEEpubid{}

\maketitle

\begin{abstract}
In recent years, digital watermarking techniques based on deep learning have been widely studied. To achieve both imperceptibility and robustness of image watermarks, most current methods employ convolutional neural networks to build robust watermarking frameworks. However, despite the success of CNN-based watermarking models, they struggle to achieve robustness against geometric attacks due to the limitations of convolutional neural networks in capturing global and long-range relationships. To address this limitation, we propose a robust watermarking framework based on the Swin Transformer, named RoWSFormer. 
Specifically, we design the Locally-Channel Enhanced Swin Transformer Block as the core of both the encoder and decoder. This block utilizes the self-attention mechanism to capture global and long-range information, thereby significantly improving adaptation to geometric distortions. Additionally, we construct the Frequency-Enhanced Transformer Block to extract frequency domain information, which further strengthens the robustness of the watermarking framework.
Experimental results demonstrate that our RoWSFormer surpasses existing state-of-the-art watermarking methods. For most non-geometric attacks, RoWSFormer improves the PSNR by 3 dB while maintaining the same extraction accuracy. In the case of geometric attacks (such as rotation, scaling, and affine transformations), RoWSFormer achieves over a 6 dB improvement in PSNR, with extraction accuracy exceeding 97\%.
\end{abstract}


\section{Introduction}
\IEEEPARstart{R}{obust} image watermarking is a technique of information hiding that is widely used for copyright protection and leakage tracing. By embedding invisible watermark message into an image, the watermark message can still be extracted even after the image has undergone severe distortion during transmission. Therefore, robust image watermarking technique possesses two key characteristics: robustness and imperceptibility. Traditional robust watermarking methods embed watermark message into the spatial domain \cite{van1994, Fridrich2001} or frequency domain \cite{KO2020,Daren2001,Urvoy2014,Fang2019screen-shooting,Chen2024Real} features of an image. However, these methods rely heavily on shallow hand-craft image features, which present some limitations in terms of robustness.

\begin{figure}[!t]
\centering
\includegraphics[width=\linewidth]{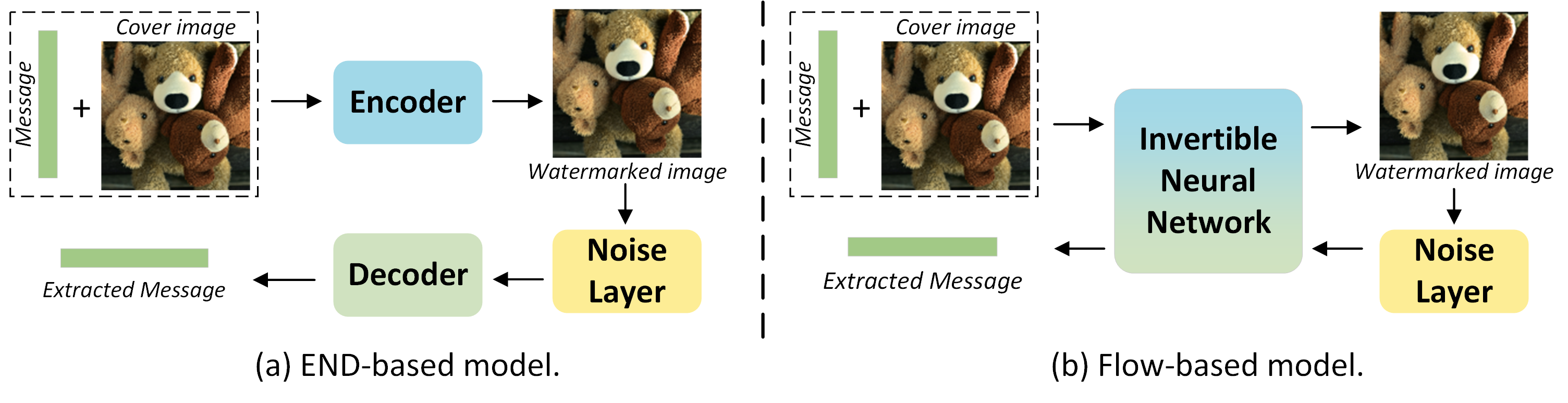}
\caption{The difference between END-based model and flow-based model.}
\label{Fig:fig1}
\end{figure}

In recent years, the rapid advancement of deep learning has led to the development of various deep-learning-based watermarking frameworks \cite{Zhu2018, Ahmadi2020, Zhang2020, Fang2022pimog, Wu2023sepmark, Fang2023denol, Guo2023practical, Qin2024Print-Camera, Liu2023wrap, Wang2024must}. These frameworks aim to address the limitations of traditional methods by fully utilizing the rich features of images, thereby enhancing watermarking robustness. Such frameworks typically consist of an encoder, a noise layer, and a decoder (END), as shown in \hyperref[Fig:fig1]{Fig. 1(a)}. The purpose of the encoder is to embed the watermark message into the cover image, while the noise layer applies distortion attacks to the watermarked image. The decoder then attempts to extract the watermark message from the attacked watermarked image. Recently, researchers have proposed another flow-based robust watermarking framework \cite{Fang2023flow}, as shown in \hyperref[Fig:fig1]{Fig. 1(b)}. This framework leverages the reversibility of Invertible Neural Networks (INNs) to enable parameter sharing between the encoding and decoding processes, thereby minimizing the embedding of redundant features. As a result, both the imperceptibility and robustness of the watermark are significantly improved.

Despite significant progress in watermarking frameworks, several critical issues persist that could compromise their effectiveness and limit their practical application in real-world scenarios. Most existing frameworks are primarily based on convolutional neural networks (CNNs). Due to the inherent limitations of convolutional operations, these CNN-based frameworks often struggle to model long-range dependencies effectively \cite{Cao2022swin}, which diminishes their ability to capture complex spatial relationships essential for robust watermarking. Furthermore, these frameworks predominantly focus on addressing typical non-geometric attacks (e.g., JPEG compression, Salt \& Pepper Noise) while neglecting common geometric attacks encountered in real-world scenarios (e.g., rotation, affine transformations). The assumption of translation invariance in CNNs further hampers their adaptability to geometric distortions. Even when geometric distortions are introduced during training through noise layer, these frameworks struggle with desynchronization issues caused by such distortions \cite{hosam2019attacking}. Additionally, the flow-based watermarking framework \cite{Fang2023flow} require the encoding and decoding networks to be completely consistent, significantly limiting the flexibility of model. This framework integrate normalizing flows using a CNN-based backbone \cite{Ye2024pprsteg}, which, due to the lack of inter-channel feature fusion, result in perceptible artifacts in the watermarked images, especially when handling robust watermarking tasks.  

In response to these limitations, researchers have begun exploring alternative approaches. Compared to CNNs, Transformers have an exceptional ability to capture global context and have been successfully applied in natural language processing (NLP) and computer vision (CV). This success has led to the development of Transformer-based watermarking framework. Recently, Lou et al. \cite{Lou2024WFormer} proposed a Transformer-based watermarking framework named WFormer. Leveraging the self-attention mechanism, WFormer effectively captures long-range dependencies in the data, extracting valid and expanded watermark features while minimizing redundancy. Additionally, WFormer incorporates a mixed attention mechanism that enables comprehensive feature fusion between the image and the watermark, achieving state-of-the-art (SOTA) performance in image watermarking tasks. However, WFormer also faces certain design limitations. Its use of fixed-size image patches may result in the loss of local and fine-grained features, restricting its ability to effectively capture multi-scale information. This limitation becomes particularly problematic in scenarios where precise spatial details are crucial for accurate watermark detection and extraction. Furthermore, the reliance of WFormer on a channel-based self-attention mechanism can lead to the loss of positional information, reducing its robustness against geometric distortions such as rotations and scaling. Consequently, WFormer struggles to address desynchronization issues caused by geometric attacks, which are common and unavoidable in real-world applications.

To address these issues, we propose a Swin Transformer-based framework for robust image watermarking, named RoWSFormer. Specifically, RoWSFormer employs the END structure to facilitate watermark embedding and extraction, enabling the encoder and decoder to operate relatively independently and thus offering greater flexibility in model design. Both the encoder and decoder utilize Locally-Channel Enhanced Swin Transformer Blocks (LCESTB) as core components to comprehensively capture channel and spatial positional information. Additionally, we have designed a Frequency-Enhanced Transformer Block (FETB) to extract frequency domain features from images, further bolstering the robustness of the watermark. Moreover, we introduce a constraint loss to regulate the encoder and prevent the generation of invalid pixel values. Extensive experimental results demonstrate that RoWSFormer surpasses current SOTAs across various attack scenarios.

The key contribution of our work can be summarised as follows:

\begin{itemize}
\item{We propose RoWSFormer, a robust image watermarking framework based on the Swin Transformer, designed to challenge and surpass the prevailing CNN-based approaches in image watermarking.}
\item{We have developed two key components for our framework: the Locally-Channel Enhanced Swin Transformer Block (LCESTB) and the Frequency-Enhanced Transformer Block (FETB). The LCESTB is designed to capture both channel and spatial positional information comprehensively, while the FETB focuses on extracting frequency domain features from images.}
\item{Extensive experiments demonstrate that our method exhibits superior performance in both visual quality and robustness compared to SOTA watermarking schemes, especially in robustness against geometric attacks.}
\end{itemize}

The remainder of this paper is organized as follows. Section II introduces the related work of Deep Learning Watermarking and Vision Transformer. Section III introduces the proposed watermarking model based on Swin Transformer. In Section IV, evaluates and analyses the results of the experiment. Section V concludes this paper.

\section{Related Work}
\subsection{Deep Learning Watermarking}
 In recent years, with the development of deep learning, many robust watermarking frameworks based on deep learning have been presented. Zhu et al.\cite{Zhu2018} first proposed the END framework HiDDeN, which successfully achieved robustness against image processing attacks (e.g., JPEG compression, Blurring) by using differentiable approximations in the noise layer to simulate certain non-differentiable noise and applying end-to-end training. Inspired by HiDDeN\cite{Zhu2018}, Tancik et al.\cite{Tancik2020stegastamp} designed StegaStamp, which addresses the robustness challenges of the print-and-capture process by mathematically simulating the printing process and generating corresponding noise layers. To incorporate real noise into the training process, Liu et al.\cite{Liu2019novel} introduced a two-stage separable deep learning network (TSDL) that effectively enhances robustness against non-differentiable and black-box attacks by fine-tuning only the decoder with real attacks in the second stage. In order to improve watermarking robustness, Fang et al.\cite{Fang2022encoded} presented an encoded feature-enhanced watermarking network based on TSDL\cite{Liu2019novel}. Nevertheless, this multi-stage training watermarking framework still lacks robustness against JPEG compression. To enhance robustness against JPEG compression, Jia et al.\cite{Jia2021mbrs} proposed a method called MBRS, which achieves high robustness against JPEG attacks by alternately using mini-batches of "real JPEG" and "simulated JPEG" noise during network training. In contrast, none of the above END watermarking models have addressed the coupling issue between the encoder and decoder. To solve this problem, Fang et al.\cite{Fang2023De-END} proposed De-END that strengthens the coupling between the encoder and decoder by using the decoder to guide the encoder. Furthermore, Fang et al.\cite{Fang2023flow} introduced a flow-based watermarking framework that leverages the reversibility of INNs to enable weight sharing between the encoder and decoder, thereby further improving robustness. Although these methods have made some progress, they still have limitations in effectively extracting and utilizing the global features of the cover image due to the inherent constraints of CNNs. To address these issues, Lou et al.\cite{Lou2024WFormer} recently proposed a Transformer-based model called WFormer, which achieves feature fusion between the image and watermark through a mixed attention mechanism. However, because WFormer uses fixed-size patches and a channel-based self-attention mechanism, it still falls short in capturing multi-scale features and positional information of the image.

\subsection{Vision Transformer}
 Vision Transformer(ViT)\cite{Dosovitskiy2020image} have demonstrated superior performance in many vision tasks, such as image classification\cite{Zhang2024Vision}, object detection\cite{Chen2022Remote}, segmentation\cite{Zheng2021rethinking,Kirillov2023segment}, and image restoration\cite{zamir2022restormer}. ViT\cite{Dosovitskiy2020image} divides images into patches (also known as tokens) and utilizes self-attention mechanisms to capture long-range dependencies between these patches. However, the quadratic computational complexity of global self-attention limits its application in high-resolution images. To address this issue, Wang et al.\cite{Wang2021pyramid} proposed the Pyramid Vision Transformer (PVT), which improves model efficiency and applicability by introducing a pyramid structure into ViT. Chen et al.\cite{Chen2021crossvit} introduced CrossViT, which uses a dual-branch structure with different-sized patches to learn multi-scale information. Liu et al.\cite{Liu2021swin} developed the Swin Transformer, which applies self-attention to local windows using a window-shifting mechanism, thus avoiding the original quadratic complexity and achieving significant results. Additionally, Wang et al.\cite{Wang2022uformer} designed UFormer for efficient image restoration, while Ke et al.\cite{Ke2024stegformer} and Lou et al.\cite{Lou2024WFormer} proposed StegFormer and WFormer, respectively, achieving efficient image steganography and SOTA performance in image watermarking. Inspired by StegFormer\cite{Ke2024stegformer}, we propose RoWSFormer, a model specifically designed for image watermarking.

\begin{figure*}[!t]
\centering
\includegraphics[width=\linewidth]{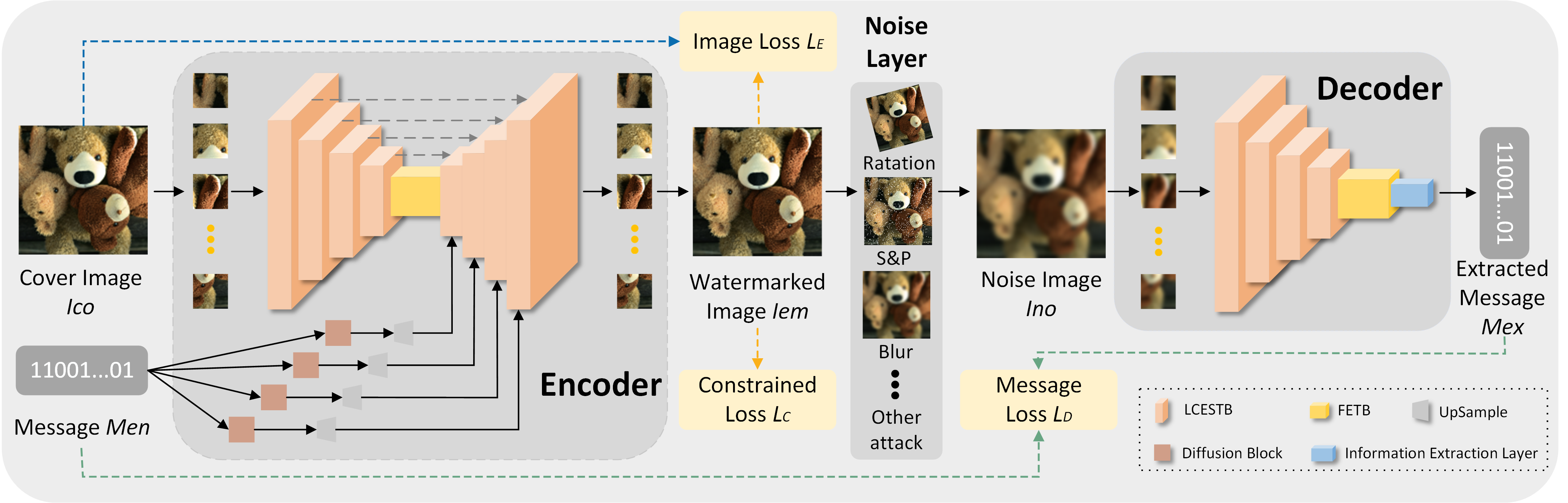}
\caption{The framework of RoWSFormer. The encoder \(E\) and decoder \(D\) consist of two crucial components: the Locally-Channel Enhanced Swin Transformer Block (LCESTB) and the Frequency-Enhanced Transformer Block (FETB). The encoder \(E\) takes the cover image \(I_{co}\) and watermark message \(M_{en}\) as input and produces the watermarked image \(I_{em}\) as output. The decoder \(D\) receives the noise image \(I_{no}\) as input and outputs the extracted watermark message \(M_{ex}\). Between the encoder and decoder is a noise layer \(N\), which includes both non-geometric and geometric distortions.}
\label{Fig:fig2}
\end{figure*}
 
\section{Our Method}
We propose an END watermarking framework based on the Swin Transformer, called \textbf{RoWSFormer}, as shown in \hyperref[Fig:fig2]{Fig. 2}. The entire framework consists of three main parts: an encoder \(E\) with parameters \(\theta_E\), a noise layer \(N\), and a decoder \(D\) with parameters \(\theta_D\). The cover image \(I_{co}\) and the watermark message \(M_{en}\) are first fed into \(E\) to generate the watermarked image \(I_{em}\). Then, the noise layer \(N\) applies attacks to \(I_{em}\) to generate the noise image \(I_{no}\), including geometric and non-geometric distortions. Lastly, \(D\) attempts to extract the watermark message \(M_{en}\) from \(I_{no}\). In the following sections, we provide a detailed description of the network architecture of the proposed framework, LCESTB and FETB.

\subsection{Model Architecture}
\subsubsection{Encoder} The primary purpose of \(E\) is to embed \(M_{en}\) into \(I_{co}\) while maintaining the visual quality of \(I_{co}\). In our proposed scheme, \(E\) adopts an architecture similar to U-Net, utilizing skip connections and multi-scale feature learning to enhance the ability of RoWSFormer to capture both the global structure and the fine details of the image, as illustrated in \hyperref[Fig:fig2]{Fig. 2}.
To be specific, given an input \( I_{co} \in \mathbb{R}^{3 \times H \times W} \), we first apply a \(3 \times 3\) convolutional layer to extract low-level features from \(I_{co}\). This operation produces an output \( I_{input} \in \mathbb{R}^{C \times H \times W} \). Next, we divide \( I_{input} \in \mathbb{R}^{C \times H \times W} \) into non-overlapping patches of size \(P \times P\) and reshape them into a flattened 2D patch sequence \( X_{token} \in \mathbb{R}^{\frac{HW}{P^2} \times P^2C} \). Here, \((H, W)\) are the dimensions of \(I_{co}\), \(C\) is the number of channels, \((P, P)\) is the size of each patch, and \(N = \frac{HW}{P^2}\) represents the total number of patches (or tokens) obtained. Following the U-Net structure, \( X_{token} \) is processed through \(K\) feature extraction stages. Each stage consists of one the proposed LCESTB and one down-sampling layer. In the down-sampling layer, we utilize a \(4 \times 4\) convolutional layer with a stride of 2, which effectively doubles the number of channels while halving the resolution of the feature maps.

Then, at the the bottleneck layer in \(E\), we incorporate one the proposed FETB. By utilizing a frequency-domain-based channel attention mechanism, the FETB effectively captures the frequency domain features of \(I_{co}\), thereby enhancing the robustness of \(I_{em}\).

Meanwhile, \(M_{en} \in \mathbb{R}^{L}\), a vector of length \(L\), is processed through several diffusion blocks. Initially, \(M_{en}\) passes through a linear layer that produces an output vector of length \(L_1\). This vector is then reshaped into a matrix of size \(L_2 \times L_2\). During the upsampling phase of \(I_{co}\) feature reconstruction, the nearest-neighbor interpolation method is used to resize \(M_{en}\) to match the dimensions of the corresponding feature maps from the downsampling stage. A \(3 \times 3\) convolutional layer is subsequently applied to increase the number of feature channels to \(C_1\).

Next, we utilize the decoder of a U-Net architecture for feature reconstruction. The reconstruction phase, much like the feature extraction process, is divided into \(K\) stages. Each reconstruction stage includes one the proposed LESTB and one upsampling layer. The upsampling layer employs a \(2 \times 2\) transposed convolution with a stride of 2, which halves the number of channels and doubles the spatial dimensions of the feature map. The upsampled feature map is then concatenated with the corresponding feature map from the feature extraction stage, along with the watermark feature map, which has been processed through the diffusion block to align with the required dimensions. This concatenated map is subsequently fed into the LESTB for image reconstruction. Ultimately, we obtain \(X_{output} \in \mathbb{R}^{\frac{HW}{P^2} \times P^2(2C + C_1)}\), which is reshaped into an image \(I_{output} \in \mathbb{R}^{(2C + C_1) \times H \times W}\). A \(3 \times 3\) convolutional layer is then applied to reduce the dimensionality of \(I_{output}\), producing the final watermarked image \(I_{em} \in \mathbb{R}^{3 \times H \times W}\).

The aim of \(E\) is generate \(I_{em}\) which approach \(I_{co}\) by updating \(\theta_E\), following the loss \(L_{E}\):
\begin{align}
L_E = \text{MSE}(I_{co}, I_{em}) = \text{MSE}(I_{co}, E(\theta_E, I_{co}, M_{en}))
\end{align}
where MSE(·) computes the mean square error.

\subsubsection{Noise Layer} The noise layer \(N\) plays a vital role in achieving robustness. Incorporating a noise layer during training can significantly enhance the robustness of the watermark\cite{hayes2017generating}. In our work, \(N\) primarily involves various types of geometric and non-geometric distortion attacks. Geometric distortion attacks include cropout, dropout, rotation, scaling, and affine attack. Non-geometric distortion attacks consist of salt-and-pepper (S\&P) noise, JPEG compression, Gaussian noise, Gaussian blur, and median blur. Since real JPEG compression is non-differentiable, we use an existing differentiable noise layer\cite{shin2017jpeg} to simulate JPEG compression.

\subsubsection{Decoder} The primary purpose of \(D\) is to extract the watermark information from \(I_{no}\). The structure of \(D\) is similar to the feature extraction process of \(E\), comprising \(K\) LESTBs, downsampling layers, a FETB, and an information extraction layer. First, we input the noise image \(I_{no} \in \mathbb{R}^{3 \times H \times W}\). We also use a \(3 \times 3\) convolution to extract the shallow features of \(I_{no}\), resulting in \(I^{input}_{no} \in \mathbb{R}^{C \times H \times W}\). \(I^{input}_{no}\) is then divided and reshaped into a sequence \(X^{de}_{token} \in \mathbb{R}^{\frac{HW}{P^2} \times P^2C}\). Using \(K\) LESTBs, downsampling layers, and an FETB, we obtain the watermark feature map of \(I_{no}\), denoted as \(I^{output}_{no} \in \mathbb{R}^{2^KC \times \frac{H}{2^K} \times \frac{W}{2^K}}\). Finally, \(I^{output}_{no}\) is passed through the information extraction layer, consisting of a convolutional layer and a fully connected layer, to extract the watermark information \(M_{ex} \in \mathbb{R}^{L}\). 

The aim of \(D\) is generate \(M_{ex}\) which approach the original watermark message \(M_{en}\) by updating \(\theta_D\), which can be formulated by:
\begin{align}
L_D = \text{MSE}(M_{en}, M_{ex}) = \text{MSE}(M_{en}, D(\theta_D, M_{en}, M_{ex}))
\end{align}

\subsection{Locally-Channel Enhanced Swin Transformer Block (LCESTB)}
Applying ViT\cite{Dosovitskiy2020image} to image watermarking tasks presents two main challenges. First, the quadratic computational cost of the self-attention mechanism renders ViTs inefficient for handling complex visual tasks. Second, previous studies\cite{li2021localvit,li2023smaller} have shown that ViT struggles to capture local and channel features, both of which are essential for effective image watermarking. While WFormer\cite{Lou2024WFormer} utilizes channel attention mechanisms to mitigate issues related to computational cost and insufficient channel feature capture, its exclusive focus on channel information limits its ability to capture local and detailed spatial information within the image. This limitation reduces its effectiveness in scenarios where precise spatial details are crucial for accurate watermark detection and extraction.

\begin{figure}[!t]
\centering
\includegraphics[width=0.9\linewidth]{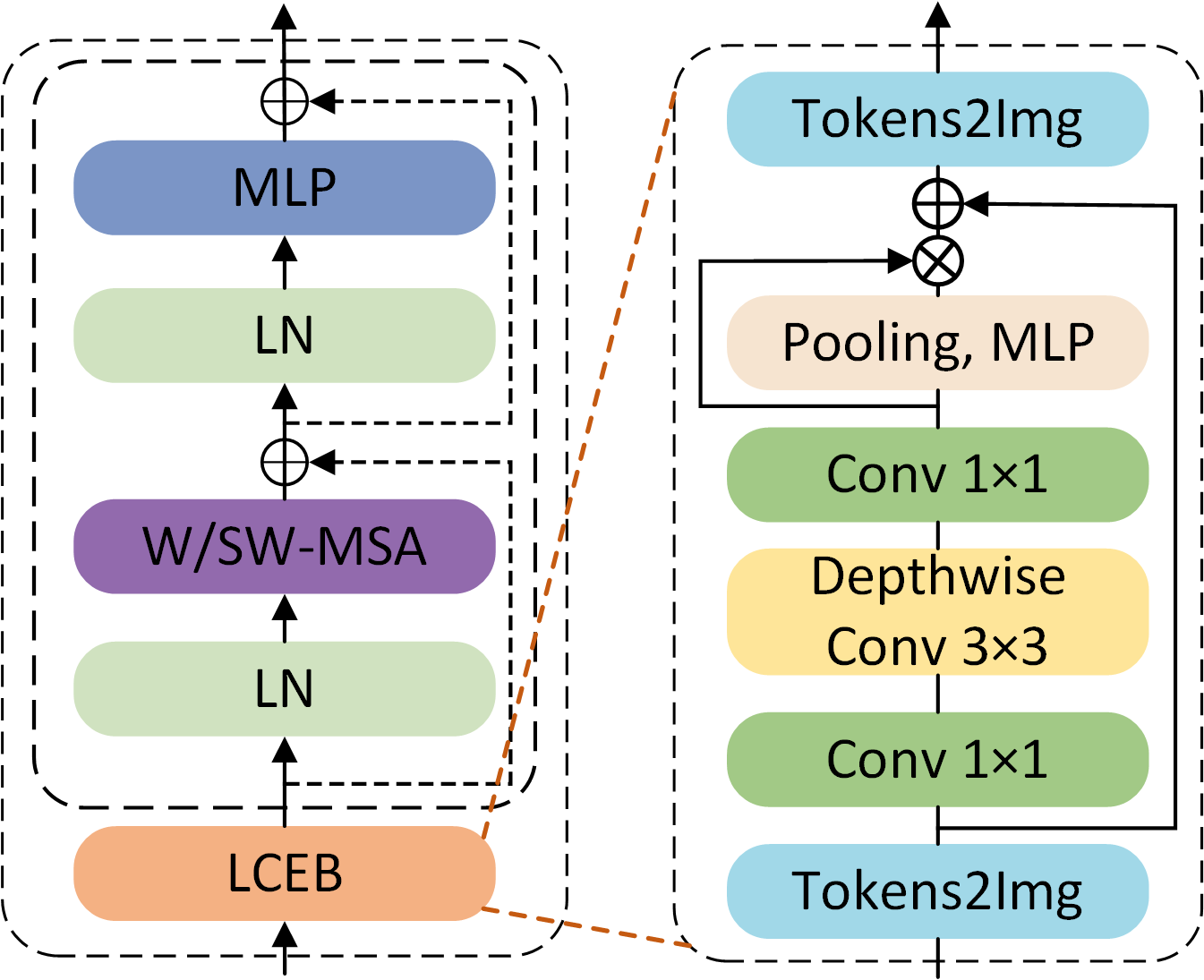}
\caption{The illustration of the Locally-Channel Enhanced Swin Transformer Block.}
\label{Fig:fig3}
\end{figure}

To address these challenges, we propose the LCESTB as a fundamental component of RoWSFormer, as illustrated in \hyperref[Fig:fig3]{Fig. 3}. Specifically, LESTB consists of two main parts. The first part is the Swin Transformer Block\cite{Liu2021swin}, which employs a window-based self-attention mechanism to effectively reduce computational costs, making the Transformer more suitable for image watermarking tasks. However, the window-based self-attention mechanism in Swin Transformer\cite{Liu2021swin} has limitations in capturing the channel features of images. To overcome this, we introduce the second part of LCESTB: the Locally-Channel Enhanced Block. This block incorporates convolution layers and channel attention mechanisms to extract both local and channel features, enhancing the capability of RoWSFormer to capture detailed channel information.

\subsubsection{Swin Transformer Block} Unlike the standard ViT\cite{Dosovitskiy2020image}, which computes self-attention over the entire image, the Swin Transformer\cite{Liu2021swin} computes self-attention within independent windows, significantly reducing computational costs. The Swin Transformer block consists of a LayerNorm (LN) layer, window-based multi-head self-attention (W-MSA), residual connections, and a 2-layer MLP with GELU nonlinearity. Due to the lack of inter-window self-attention in W-MSA, the Swin Transformer uses two consecutive Swin Transformer blocks and a shifted window approach to establish connections between windows. This consecutive Swin Transformer block\cite{Liu2021swin} with shifted windows can be represented as:
\begin{align}
    \hat{X}^l &= \text{W-MSA}(\text{LN}(X^{l-1})) + X^{l-1}, \notag \\
    X^l &= \text{MLP}(\text{LN}(\hat{X}^l)) + \hat{X}^l, \notag \\
    \hat{X}^{l+1} &= \text{SW-MSA}(\text{LN}(X^l)) + X^l, \notag \\
    X^{l+1} &= \text{MLP}(\text{LN}(\hat{X}^{l+1})) + \hat{X}^{l+1}
\end{align}
where \(\hat{X}^l\) and \(X^l\) denote the output features of the (S)W-MSA module and the MLP module for block \(l\), respectively.

\subsubsection{Locally-Channel Enhanced Block} Transformers tend to focus on global modeling, which limits their ability to capture local features and channel information. To improve the performance of RoWSFormer, we propose the Locally-Channel Enhanced Block to effectively extract both local features and channel information from images.

As shown in \hyperref[Fig:fig3]{Fig. 3}, we first reshape \( X_{token} \in \mathbb{R}^{\frac{HW}{P^2} \times P^2C} \) into an image \( X_{img} \in \mathbb{R}^{C \times H \times W} \). A linear projection layer is applied to increase the dimensionality, allowing for better capture of channel information. This is followed by a \(3 \times 3\) depthwise convolution to capture local features. Afterward, we apply another linear projection layer to reduce the number of channels, aligning the output with the input dimensions and resulting in \( X_{channel} \in \mathbb{R}^{C \times H \times W} \). Next, a pooling layer followed by a fully connected layer is used to compute attention weights for each channel, resulting in \( X_{weight} \in \mathbb{R}^C \). We multiply \( X_{channel} \) by \( X_{weight} \) to generate a bias \( X_{bias} \in \mathbb{R}^{C \times H \times W} \) that incorporates both local features and channel information. Finally, we add \( X_{bias} \) to \( X_{img} \), reshape the result back into tokens, and use it as input for the Swin Transformer block.

\subsection{Frequency-Enhanced Transformer Block (FETB)} It is important to note that although the Swin Transformer\cite{Liu2021swin} employs a shifted window approach to establish connections between windows, its window-based self-attention still constrains the ability of Swin Transformer\cite{Liu2021swin} to capture global features. Given that the image undergoes multiple downsampling operations, the size of the feature map is significantly reduced. As a result, even with global self-attention, the computational cost remains relatively low. Therefore, in FETB, we use a standard ViT\cite{Dosovitskiy2020image} instead of the Swin Transformer\cite{Liu2021swin} to achieve global modeling.

\begin{figure}[!t]
\centering
\includegraphics[width=0.9\linewidth]{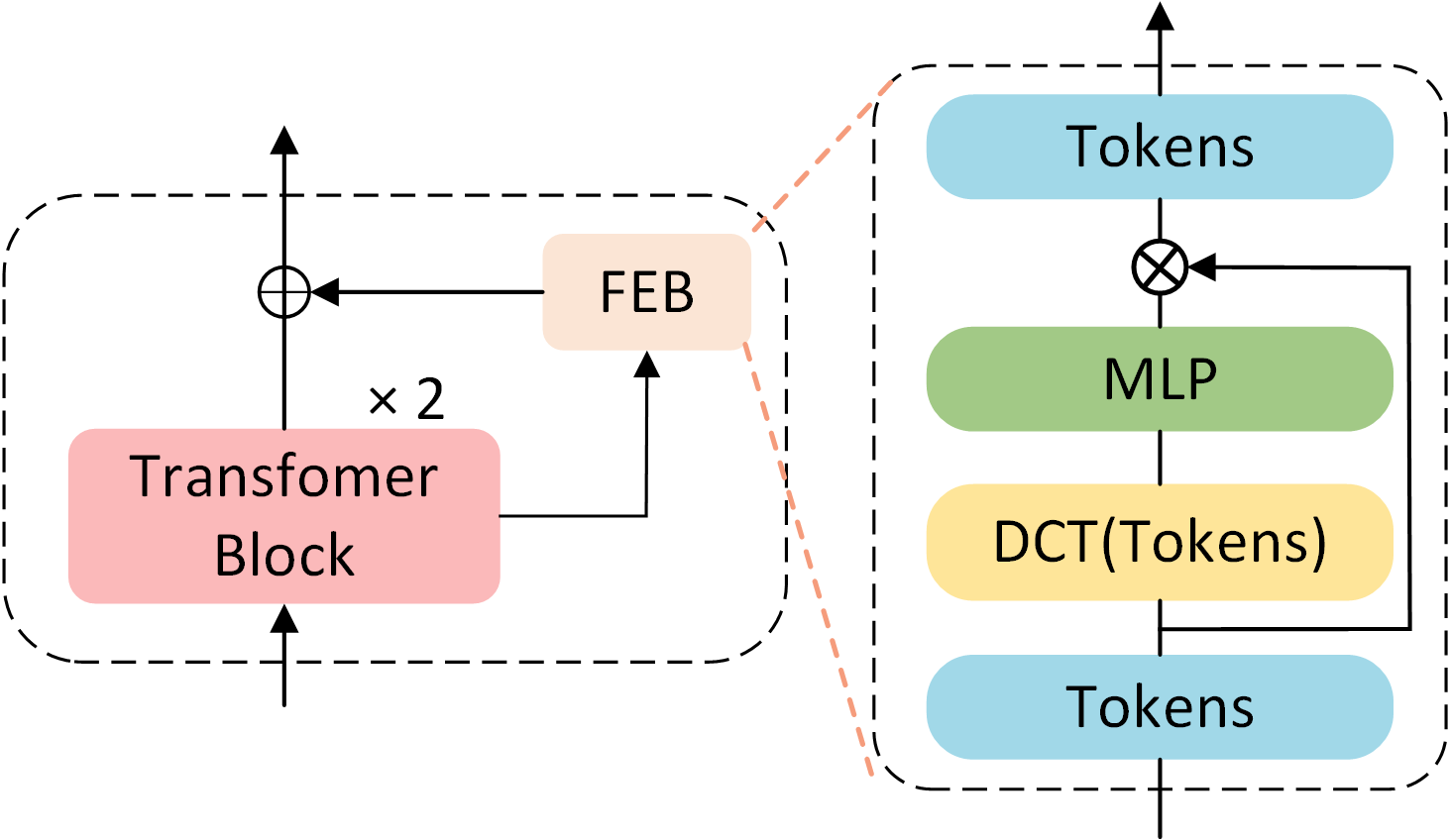}
\caption{The illustration of the Frequency-Enhanced Transformer Block.}
\label{Fig:fig4}
\end{figure}

Moreover, network models often struggle to capture the rich frequency information inherent in real-world datasets\cite{JIANG2023FECAM}. Research\cite{CAO2024Universal} has shown that, in watermarking networks, the frequency differences between the watermark mask and the carrier image can greatly influence performance. Thus, it is essential for the network to learn richer frequency domain features. Inspired by this, we introduce the Frequency Enhance Transformer Block (FETB) to improve the performance of RoWSFormer, as shown in \hyperref[Fig:fig4]{Fig. 4}. Specifically, the FETB is composed of multiple standard Transformer blocks\cite{Dosovitskiy2020image} and a frequency enhancement block.

\subsubsection{Transformer Block} We employ multiple ViT Blocks\cite{Dosovitskiy2020image} to effectively capture global features across the entire image. The process is as follows:
\begin{align}
    \hat{X}^l &= \text{MSA}(\text{LN}(X^{l-1})) + X^{l-1}, \notag \\
    X^l &= \text{MLP}(\text{LN}(\hat{X}^l)) + \hat{X}^l,
\end{align}
where $\hat{X}^l$ and $X^l$ denote the output features of the MSA module and the MLP module for block $l$, respectively.

\subsubsection{Frequency-Enhanced Block} After processing through the Transformer, and inspired by FECAM\cite{JIANG2023FECAM}, we designed the Frequency-Enhanced Block to extract frequency domain features from images, as shown in \hyperref[Fig:fig4]{Fig. 4}. First, we divide the output of the Transformer \(X^T_{token} \in \mathbb{R}^{\frac{HW}{P^2} \times P^2C}\) into \(P^2C\) groups along the channel dimension. Then, each group undergoes a Discrete Cosine Transform (DCT), followed by a stacking operation to obtain the DCT frequency domain attention vector \( X_{Freq} \in \mathbb{R}^{\frac{HW}{P^2} \times P^2C} \). Finally, a simple fully connected(FC) layer is used to compute the frequency domain attention weights \( F_w \in \mathbb{R}^{P^2C} \). The output of the Frequency Enhance Block \(X_{output}\) is obtained by multiplying \( F_w \) with \( X^T_{token} \).

The entire process can be expressed mathematically as:
\begin{align}
\label{deqn_ex1a}
F_w &= \text{FC}(\text{stack}(\text{DCT}(X^T_{token}))), \notag \\
X_{output} &= F_w \cdot X^T_{token},
\end{align}

\subsection{Loss Function}
Due to the watermark embedding process, some pixel values in \(I_{co}\) may fall outside the standard range of [0, 255]. To address this issue, we propose a constrained loss function \(L_{C}\), which encourages \(E\) to ensure that the pixel values of \(I_{em}\) remain within the [0, 255] range. \(L_{C}\) is defined as follows:
\begin{align}
L_{C} = \sum_{i=1}^{H} \sum_{j=1}^{W} \begin{cases}
\frac{1}{2}\lvert I_{em}(i, j) - 1 \rvert, & \text{if } I_{em}(i, j) > 1, \\
\frac{1}{2}\lvert I_{em}(i, j) \rvert, & \text{if } I_{em}(i, j) < 0, \\
0, & \text{otherwise}.
\end{cases}
\end{align}

So, the total loss function \(L_{total}\) is consist of image loss, decoding loss and constrained loss, which can be formulated by:
\begin{align}
{L_{total}} = \lambda_1 {L}_{E} + \lambda_2 {L}_{D} + \lambda_3 {L}_{C}
\end{align}
where \(\lambda_1\), \(\lambda_2\), and \(\lambda_3\) are weight factors.

\begin{figure*}[!t]
\centering
\includegraphics[width=0.9\linewidth]{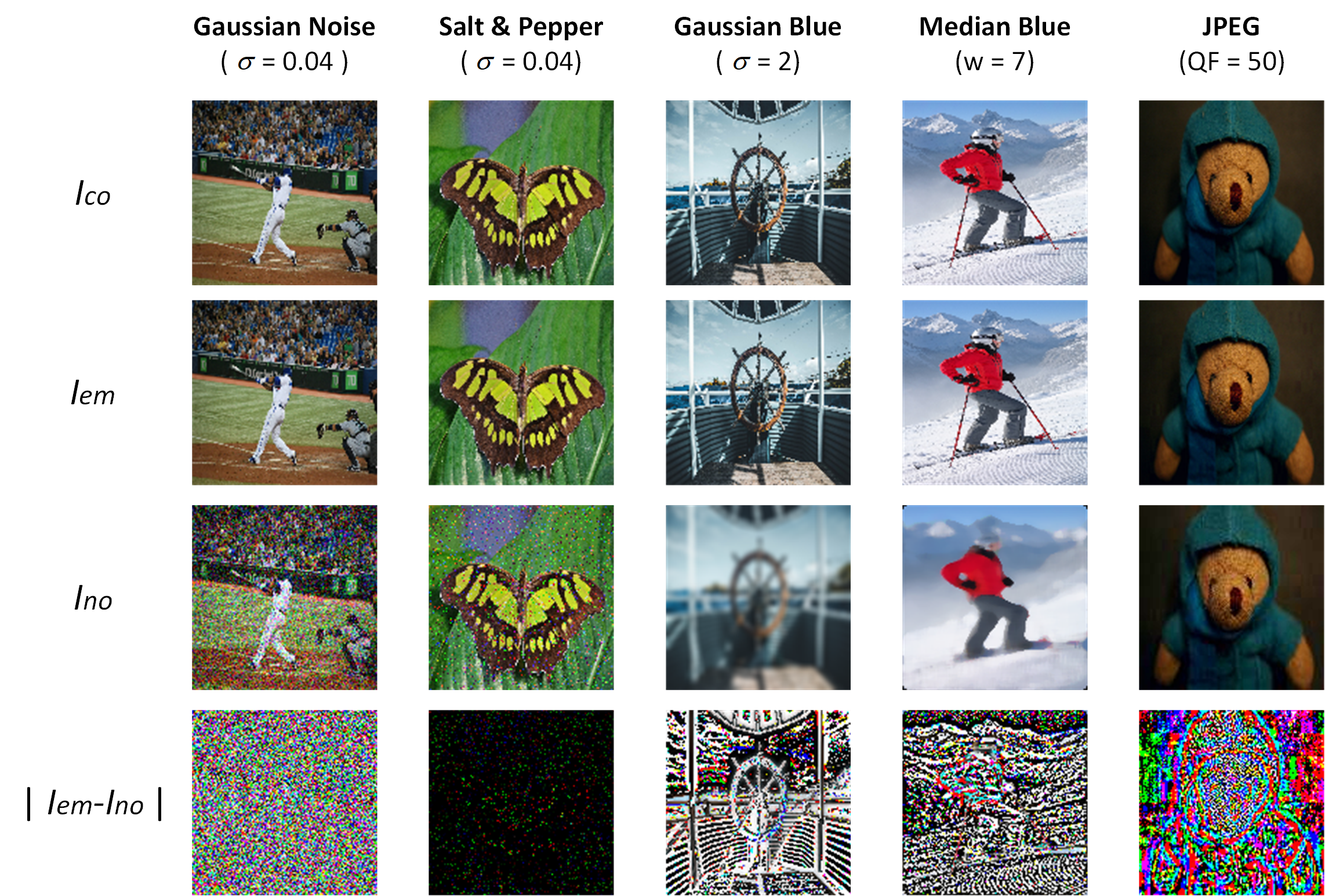}
\caption{The watermarked image and the corresponding image with non-geometric distortions. Top: the cover image \(I_{co}\); Second: the encoded image \(I_{em}\); Third: the noise image \(I_{no}\); Bottom: the residual image $|$\(I_{em}\)$-$\(I_{no}\)$|$.}
\label{Fig:fig5}
\end{figure*}

\section{Experimental Evaluation}
\subsection{Experimental Settings}
\subsubsection{Basic Settings} Our RoWSFormer model is implemented using PyTorch \cite{Collobert2011torch7} and executed on an NVIDIA GeForce RTX 4090. To maintain consistency with other methods, all images are resized to 128 $\times$ 128, and the watermark length \(L\) is set to 64. The parameters \(\lambda_1\), \(\lambda_2\), and \(\lambda_3\) are fixed at 2, 10, and 0.1, respectively. The AdamW optimizer is employed to train RoWSFormer, using a cosine decay strategy to gradually reduce the learning rate from an initial value of 1e-3 to 1e-6.

\subsubsection{Datasets} The DIV2K\cite{agustsson2017ntire} dataset is used to train our RoWSFormer model. To evaluate its generalization ability, we use both the DIV2K\cite{agustsson2017ntire} and COCO\cite{lin2014microsoft} datasets. Specifically, for the COCO\cite{lin2014microsoft} dataset, we randomly select 5,000 images to serve as the test set.

\subsubsection{Benchmarks} To demonstrate the invisibility and robustness of the proposed RoWSFormer, we compare it against several SOTA watermarking methods, including three CNN-based methods: HiDDeN\cite{Zhu2018}, TSDL\cite{Liu2019novel}, and MBRS\cite{Jia2021mbrs}; a normalizing flow-based method: FBRW\cite{Fang2023flow}; and a Transformer-based method: WFormer\cite{Lou2024WFormer}. For robustness testing, we use five non-geometric distortions ("Gaussian Noise", "Salt \& Pepper Noise", "Gaussian Blur", "Median Blur", and "JPEG Compression") and five geometric distortions ("Cropout", "Dropout", "Rotation", "scaling", and "Affine Attack"). To accurately assess robustness, we train a specific watermarking network for each type of distortion. For a fair comparison, all watermarking methods are retrained using the same dataset and noise layer. All compared experiments are conducted on images with a size of 128 $\times$ 128, and the watermark length is set to 64.

\subsubsection{Metrics} In this paper, we use Peak Signal-to-Noise Ratio (PSNR) to evaluate the imperceptibility of the watermark, with higher values indicating better imperceptibility. Additionally, we use extraction bit accuracy (ACC) to assess the robustness of the proposed model, with higher ACC values reflecting greater robustness.

\subsection{Invisibility And Robustness Against Non-geometric Attacks}
In this section, we assess the invisibility and robustness of our method, along with SOTAs, against non-geometric attacks. We conduct experiments using various types of noise, as shown in \hyperref[Fig:fig5]{Fig. 5}.

\subsubsection{Gaussian Noise} Gaussian noise, which follows a Gaussian distribution, is frequently encountered in message transmission. In our experiments, we introduce Gaussian noise with variances ranging from 0.001 to 0.04 during the training phase, and adjust the variance between 0.01 and 0.05 during the testing phase. The results of these experiments are presented in \hyperref[table1]{Table I}.

\begin{table}[ht]
\centering
\caption{PSNR and ACC with Different Ratio of Gaussian Noise}
\label{table1}
\begin{tabular}{>{\centering\arraybackslash}p{0.091\textwidth} >{\centering\arraybackslash}p{0.055\textwidth} >{\centering\arraybackslash}p{0.035\textwidth} >{\centering\arraybackslash}p{0.03\textwidth} >{\centering\arraybackslash}p{0.03\textwidth} >{\centering\arraybackslash}p{0.03\textwidth} >{\centering\arraybackslash}p{0.03\textwidth}}
\cmidrule(lr){1-7}
\multirow{2}{*}{Model} & \centering \multirow{2}{*}{PSNR(dB)} & \multicolumn{5}{c}{ACC(\%)} \\
\cmidrule(lr){3-7}
 & & $\sigma$=0.01 & 0.02 & 0.03 & 0.04 & 0.05 \\
\cmidrule(lr){1-7}
HiDDeN \cite{Zhu2018} & 36.25 & 89.58 & 86.46 & 83.96 & 83.12 & 79.17 \\
TSDL \cite{Liu2019novel} & 39.46 & 92.08 & 91.25 & 88.33 & 87.08 & 82.92 \\
MBRS \cite{Jia2021mbrs} & 39.70 & 99.91 & 99.42 & 98.10 & 96.09 & 94.15 \\
FBRW \cite{Fang2023flow} & 40.05 & \textbf{100} & \textbf{99.98} & \textbf{99.94} & \textbf{99.89} & \textbf{98.83} \\
WFormer \cite{Lou2024WFormer} & \textbf{40.48} & \textbf{100} & \textbf{99.98} & 99.83 & 99.41 & 98.72 \\
Ours & 39.87 & \textbf{100} & 99.92 & 99.48 & 98.32 & 97.29 \\
\cmidrule(lr){1-7}
\end{tabular}
\end{table}

Although our model does not achieve SOTA performance in defending against Gaussian noise attacks, it still demonstrates strong imperceptibility and robustness. The PSNR reaches close to 40, and the extraction accuracy is within 2 percentage points of the SOTA model. This indicates that, despite not being the top performer, our model remains highly effective in handling Gaussian noise.

\subsubsection{Salt \& Pepper Noise} Salt \& Pepper noise, like Gaussian noise, is commonly encountered in transmission processes, where a certain percentage of image pixels is randomly corrupted. In the training phase, we apply Salt \& Pepper noise by selecting a random ratio between 0.001 and 0.04. During testing, the noise ratio is adjusted between 0.01 and 0.05 to evaluate the performance of different models under varying noise levels. The final results are presented in \hyperref[table2]{Table II}.

\begin{table}[ht]
\centering
\caption{PSNR and ACC with Different Ratio of Salt \& Pepper Noise}
\label{table2}
\begin{tabular}{>{\centering\arraybackslash}p{0.091\textwidth} >{\centering\arraybackslash}p{0.055\textwidth} >{\centering\arraybackslash}p{0.035\textwidth} >{\centering\arraybackslash}p{0.03\textwidth} >{\centering\arraybackslash}p{0.03\textwidth} >{\centering\arraybackslash}p{0.03\textwidth} >{\centering\arraybackslash}p{0.03\textwidth}}
\cmidrule(lr){1-7}
\multirow{2}{*}{Model} & \centering \multirow{2}{*}{PSNR(dB)} & \multicolumn{5}{c}{ACC(\%)} \\
\cmidrule(lr){3-7}
 & & $\sigma$=0.01 & 0.02 & 0.03 & 0.04 & 0.05 \\
\cmidrule(lr){1-7}
HiDDeN \cite{Zhu2018} & 46.04 & 95.12 & 93.79 & 93.45 & 92.92 & 90.42 \\
TSDL \cite{Liu2019novel} & 51.16 & 97.29 & 95.63 & 93.54 & 92.71 & 91.46 \\
MBRS \cite{Jia2021mbrs} & 51.79 & 98.05 & 98.74 & 98.34 & 97.56 & 96.68 \\
FBRW \cite{Fang2023flow} & 51.97 & \textbf{100} & \textbf{100} & \textbf{100} & \textbf{100} & \textbf{100} \\
WFormer \cite{Lou2024WFormer} & 52.71 & 99.90 & 99.83 & 99.74 & 99.52 & 99.22 \\
Ours & \textbf{55.76} & \textbf{100} & \textbf{100} & 99.98 & 99.96 & 99.98 \\
\cmidrule(lr){1-7}
\end{tabular}
\end{table}

Compared to other methods, our proposed RoWSFormer delivers superior performance, achieving a PSNR value exceeding 55 dB. It also demonstrates outstanding robustness, with extraction accuracy surpassing 99\% across all tested Salt \& Pepper Noise ratios. This highlights RoWSFormer’s exceptional resilience to Salt \& Pepper noise.

\subsubsection{Gaussian Blur} For Gaussian blur distortion, we set a fixed variance of 2 for the noise layer during the training phase. In the testing phase, we vary the variance of the Gaussian blur from 0.0001 to 2 assess both the imperceptibility and robustness of methods. A detailed comparison of visual quality and extraction accuracy under these conditions is provided in \hyperref[table3]{Table III}.

\begin{table}[ht]
\centering
\caption{PSNR and ACC with Different Ratio of Gaussian Blur}
\label{table3}
\begin{tabular}{>{\centering\arraybackslash}p{0.091\textwidth} >{\centering\arraybackslash}p{0.055\textwidth} >{\centering\arraybackslash}p{0.055\textwidth} >{\centering\arraybackslash}p{0.04\textwidth} >{\centering\arraybackslash}p{0.04\textwidth} >{\centering\arraybackslash}p{0.04\textwidth} }
\cmidrule(lr){1-6}
\multirow{2}{*}{Model} & \centering \multirow{2}{*}{PSNR(dB)} & \multicolumn{4}{c}{ACC(\%)} \\
\cmidrule(lr){3-6}
 & & $\sigma$=0.0001 & 0.5 & 1 & 2 \\
\cmidrule(lr){1-6}
HiDDeN \cite{Zhu2018} & 46.21 & 95.44 & 95.21 & 94.33 & 84.37 \\
TSDL \cite{Liu2019novel} & 45.07 & 99.92 & 99.79 & 98.48 & 93.21 \\
MBRS \cite{Jia2021mbrs} & 47.91 & 98.64 & 98.25 & 97.66 & 87.80 \\
FBRW \cite{Fang2023flow} & 48.09 & 99.97 & 99.87 & 99.65 & 98.16 \\
WFormer \cite{Lou2024WFormer} & 49.36 & 98.90 & 98.96 & 99.01 & 98.69 \\
Ours & \textbf{52.45} & \textbf{100} & \textbf{100} & \textbf{100} & \textbf{100} \\ 
\cmidrule(lr){1-6}
\end{tabular}
\end{table}

The results show that RoWSFormer delivers exceptional performance, achieving a PSNR of over 52 dB for watermarked images, significantly outperforming other methods in terms of visual quality. Moreover, RoWSFormer improves extraction accuracy by 1\% compared to competing schemes, further demonstrating its superior robustness against Gaussian blur distortion.

\subsubsection{Median Blur} Median Blur is a widely used technique in image processing, particularly for reducing noise. To ensure robustness during training, we apply a fixed blurring window size of 7 $\times$ 7. In the testing phase, we evaluate robustness using varying window sizes of 3 $\times$ 3, 5 $\times$ 5, and 7 $\times$ 7. The results of these experiments are presented in \hyperref[table4]{Table IV}.

\begin{table}[ht]
\centering
\caption{PSNR and ACC with Different Windows of Median Blur}
\label{table4}
\begin{tabular}{>{\centering\arraybackslash}p{0.091\textwidth} >{\centering\arraybackslash}p{0.055\textwidth} >{\centering\arraybackslash}p{0.07\textwidth} >{\centering\arraybackslash}p{0.05\textwidth} >{\centering\arraybackslash}p{0.05\textwidth} }
\cmidrule(lr){1-5}
\multirow{2}{*}{Model} & \centering \multirow{2}{*}{PSNR(dB)} & \multicolumn{3}{c}{ACC(\%)} \\
\cmidrule(lr){3-5}
 & & $w$=3 $\times$ 3 & 5 $\times$ 5 & 7 $\times$ 7 \\
\cmidrule(lr){1-5}
HiDDeN \cite{Zhu2018} & 37.07 & 86.25 & 83.70 & 79.71 \\
TSDL \cite{Liu2019novel} & 38.64 & 99.38 & 97.21 & 95.12 \\
MBRS \cite{Jia2021mbrs} & 40.98 & 99.42 & 98.93 & 97.27 \\
FBRW \cite{Fang2023flow} & 41.47 & \textbf{100} & \textbf{100} & \textbf{100} \\
WFormer \cite{Lou2024WFormer} & 44.76 & 99.93 & 99.85 & 99.55 \\
Ours & \textbf{48.27} & 99.31 & 99.26 & 98.53 \\ 
\cmidrule(lr){1-5}
\end{tabular}
\end{table}

RoWSFormer demonstrates impressive performance, achieving a PSNR of 48.27 dB under Median Blur, outperforming other methods by approximately 3 dB. While it may not be the most robust model against Median Blur, it still maintains an extraction accuracy above 98\%, showcasing its strong resilience in handling such distortions.

\subsubsection{JPEG Compression} JPEG compression is commonly encountered during image saving and format conversion. In the training stage, we configure the noise layer with a quality factor (QF) of 50. To assess the model's robustness against JPEG compression attacks, we conduct tests with QF values ranging from 40 to 90. The results of these experiments are detailed in \hyperref[table5]{Table V}.

\begin{table}[ht]
\centering
\caption{PSNR and ACC with Different QF of JPEG}
\label{table5}
\begin{tabular}{>{\centering\arraybackslash}p{0.091\textwidth} >{\centering\arraybackslash}p{0.045\textwidth} >{\centering\arraybackslash}p{0.03\textwidth} >{\centering\arraybackslash}p{0.025\textwidth} >{\centering\arraybackslash}p{0.025\textwidth} >{\centering\arraybackslash}p{0.025\textwidth} >{\centering\arraybackslash}p{0.025\textwidth} >{\centering\arraybackslash}p{0.025\textwidth} }
\cmidrule(lr){1-8}
\multirow{2}{*}{Model} & \centering \multirow{2}{*}{PSNR(dB)} & \multicolumn{6}{c}{ACC(\%)} \\
\cmidrule(lr){3-8}
 & & QF=40 & 50 & 60 & 70 & 80 & 90 \\
\cmidrule(lr){1-8}
HiDDeN \cite{Zhu2018} & 33.29 & 86.67 & 91.24 & 92.92 & 93.33 & 93.54 & 94.38 \\
TSDL \cite{Liu2019novel} & 39.39 & 91.04 & 91.46 & 93.96 & 94,21 & 94,35 & 94.74 \\
MBRS \cite{Jia2021mbrs} & 45.16 & 94.83 & 94.93 & 96.68 & 97.66 & 97.66 & 98.84 \\
FBRW \cite{Fang2023flow} & \textbf{47.21} & \textbf{99.71} & \textbf{100} & \textbf{100} & \textbf{100} & \textbf{100} & \textbf{100} \\
WFormer \cite{Lou2024WFormer} & 45.41 & 95.83 & 98.79 & 99.60 & 99.92 & 99.96 & \textbf{100} \\
Ours & 45.04 & 96.15 & 97.68 & 98.49 & 99.61 & 99.99 & \textbf{100} \\
\cmidrule(lr){1-8}
\end{tabular}
\end{table}

While FBRW \cite{Fang2023flow} achieves the highest PSNR and extraction accuracy across various QF values for JPEG compression, our proposed model also delivers excellent image quality and exhibits strong robustness.

\subsection{Invisibility And Robustness Against Geometric Attacks}
\begin{figure*}[!t]
\centering
\includegraphics[width=0.9\linewidth]{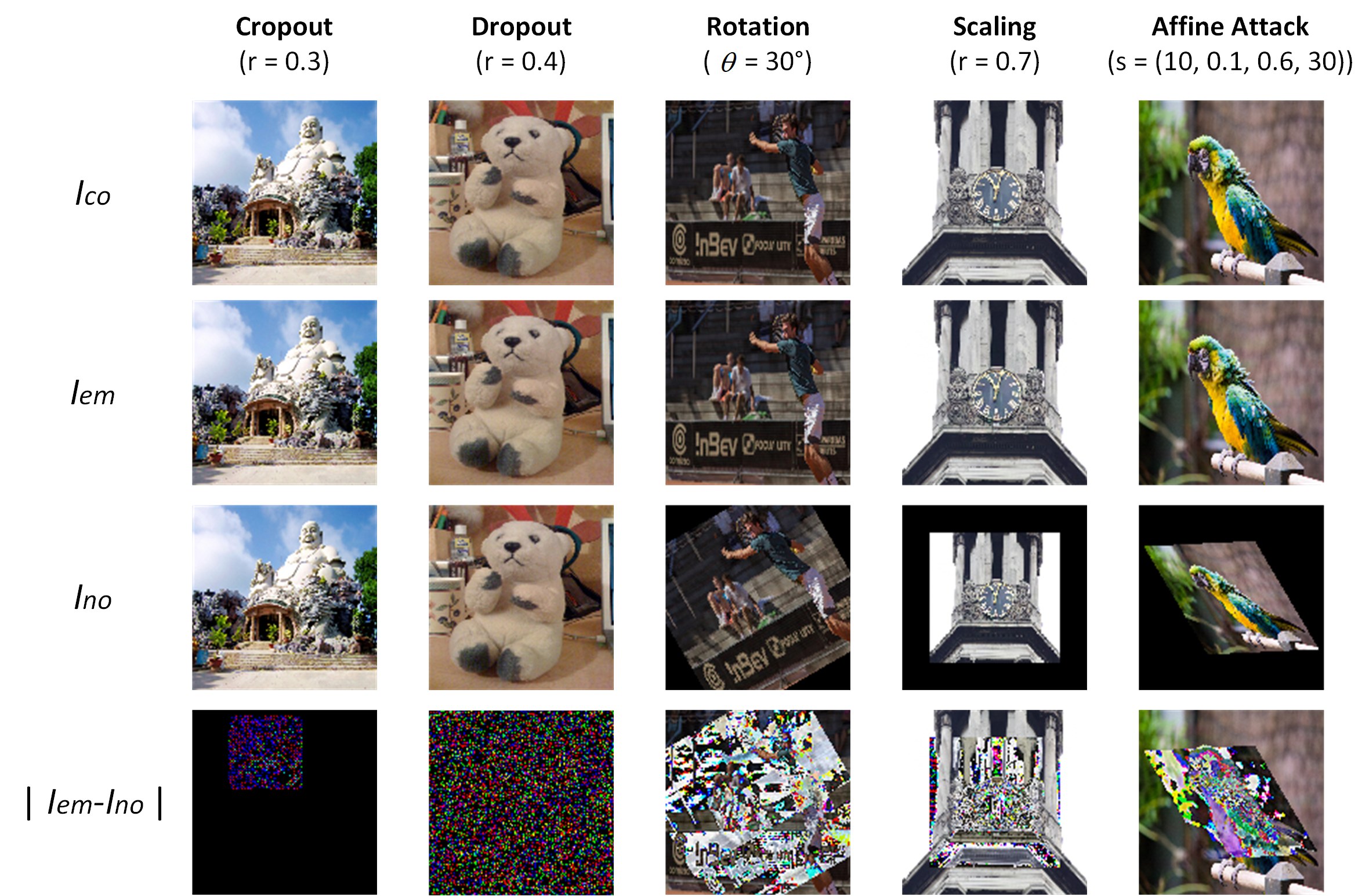}
\caption{The watermarked image and the corresponding image with geometric distortions.}
\label{Fig:fig6}
\end{figure*}

In this section, we evaluate the invisibility and robustness of our method, along with SOTA models, in the face of geometric attacks. We conduct experiments with various types of noise, as illustrated in \hyperref[Fig:fig6]{Fig. 6}.

\subsubsection{Cropout} Cropout is a type of distortion where part of the watermarked image is preserved, while the remaining area is replaced with the corresponding region from the original image. During the training stage, we apply a cropout ratio of 0.4. In the testing phase, we evaluate the effect of different models by varying the cropout ratio from 0.1 to 0.5. The experimental results are provided in \hyperref[table6]{Table VI}.

\begin{table}[ht]
\centering
\caption{PSNR and ACC with Different Ratio of Cropout}
\label{table6}
\begin{tabular}{>{\centering\arraybackslash}p{0.091\textwidth} >{\centering\arraybackslash}p{0.045\textwidth} >{\centering\arraybackslash}p{0.035\textwidth} >{\centering\arraybackslash}p{0.03\textwidth} >{\centering\arraybackslash}p{0.03\textwidth} >{\centering\arraybackslash}p{0.03\textwidth} >{\centering\arraybackslash}p{0.03\textwidth}}
\cmidrule(lr){1-7}
\multirow{2}{*}{Model} & \centering \multirow{2}{*}{PSNR(dB)} & \multicolumn{5}{c}{ACC(\%)} \\
\cmidrule(lr){3-7}
 & & $r$=0.1 & 0.2 & 0.3 & 0.4 & 0.5 \\
\cmidrule(lr){1-7}
HiDDeN \cite{Zhu2018} & 40.62 & 95.63 & 94.73 & 88.75 & 76.88 & 61.67 \\
TSDL \cite{Liu2019novel} & 47.48 & 98.72 & 98.54 & 96.88 & 93.75 & 93.21 \\
MBRS \cite{Jia2021mbrs} & 48.05 & 99.71 & 99.22 & 97.18 & 90.43 & 83.50 \\
WFormer \cite{Lou2024WFormer} & 52.72 & 99.99 & 99.98 & 99.97 & 99.96 & 98.35 \\
Ours & \textbf{54.74} & \textbf{100} & \textbf{100} & \textbf{99.99} & \textbf{99.98} & \textbf{99.23} \\
\cmidrule(lr){1-7}
\end{tabular}
\end{table}

The proposed method achieves an impressive PSNR of 54.74 for watermarked images, ensuring top-tier visual quality. Despite this high level of image quality, the method also demonstrates remarkable robustness across different cropout ratios. Specifically, for cropout ratios ranging from 0.1 to 0.5, the extraction accuracy consistently exceeds 99\%. As the cropout ratio increases, the advantage of the proposed model become even more pronounced.

\subsubsection{Dropout} Dropout distortion involves randomly replacing a certain percentage of image pixels with pixels from the original image, differing from cropout where the replacement occurs in a specific region. In dropout, the replacement pixels are distributed randomly across the entire image. For training, we use a dropout ratio of 0.4. During testing, we vary this ratio from 0.2 to 0.6 to assess performance under different levels of distortion, as detailed in \hyperref[table7]{Table VII}. 

\begin{table}[ht]
\centering
\caption{PSNR and ACC with Different Ratio of Dropout}
\label{table7}
\begin{tabular}{>{\centering\arraybackslash}p{0.091\textwidth} >{\centering\arraybackslash}p{0.057\textwidth} >{\centering\arraybackslash}p{0.035\textwidth} >{\centering\arraybackslash}p{0.03\textwidth} >{\centering\arraybackslash}p{0.03\textwidth} >{\centering\arraybackslash}p{0.03\textwidth} >{\centering\arraybackslash}p{0.03\textwidth}}
\cmidrule(lr){1-7}
\multirow{2}{*}{Model} & \centering \multirow{2}{*}{PSNR(dB)} & \multicolumn{5}{c}{ACC(\%)} \\
\cmidrule(lr){3-7}
 & & $r$=0.2 & 0.3 & 0.4 & 0.5 & 0.6 \\
\cmidrule(lr){1-7}
HiDDeN \cite{Zhu2018} & 42.59 & 90.21 & 89.58 & 87.08 & 86.74 & 82.71 \\
TSDL \cite{Liu2019novel} & 53.59 & 97.54 & 95.21 & 93.54 & 92.29 & 90.42 \\
MBRS \cite{Jia2021mbrs} & 58.63 & 96.29 & 94.73 & 94.15 & 92.58 & 90.63 \\
WFormer \cite{Lou2024WFormer} & 58.99 & 99.59 & 99.22 & 98.68 & 97.66 & 95.84 \\
Ours & \textbf{61.43} & \textbf{100} & \textbf{100} & \textbf{100} & \textbf{99.98} & \textbf{99.16} \\
\cmidrule(lr){1-7}
\end{tabular}
\end{table}

The proposed RoWSFormer not only achieves the highest visual quality among all methods but also consistently outperforms other frameworks in terms of robustness against dropout distortion.

\subsubsection{Rotation} Rotation distortion involves rotating an image by a specific angle, which can significantly affect the alignment of embedded watermarks and challenge their accurate extraction. To enhance robustness against rotational variations, we randomly select rotation angles between $-30^\circ$ and $30^\circ$ during the training phase. In the testing phase, we assess the model's performance by applying rotation angles within the same range of $-30^\circ$ to $30^\circ$. The experimental results for this type of rotation distortion are presented in \hyperref[table8]{Table VIII}.

\begin{table}[ht]
\centering
\caption{PSNR and ACC with Different Angle of Rotation}
\label{table8}
\begin{tabular}{>{\centering\arraybackslash}p{0.091\textwidth} >{\centering\arraybackslash}p{0.057\textwidth} >{\centering\arraybackslash}p{0.068\textwidth} >{\centering\arraybackslash}p{0.025\textwidth} >{\centering\arraybackslash}p{0.025\textwidth} >{\centering\arraybackslash}p{0.025\textwidth} >{\centering\arraybackslash}p{0.025\textwidth}}
\cmidrule(lr){1-7}
\multirow{2}{*}{Model} & \centering \multirow{2}{*}{PSNR(dB)} & \multicolumn{5}{c}{ACC(\%)} \\
\cmidrule(lr){3-7}
 & & $\theta=-30^\circ$ & $-15^\circ$ & $0^\circ$ & $15^\circ$ & $30^\circ$ \\
\cmidrule(lr){1-7}
HiDDeN \cite{Zhu2018} & 36.35 & 92.68 & 93.13 & 97.76 & 93.27 & 92.04  \\
TSDL \cite{Liu2019novel} & 36.22 & 87.54 & 91.57 & 95.27 & 91.64 & 87.57  \\
MBRS \cite{Jia2021mbrs} & 35.58 & 81.88 & 88.69 & 92.42 & 88.76 & 81.74 \\
WFormer \cite{Lou2024WFormer} & 43.79 & 94.64 & 97.83 & \textbf{100} & 97.96 & 94.77 \\
Ours & \textbf{50.26} & \textbf{99.97} & \textbf{99.99} & \textbf{100} & \textbf{100} & \textbf{99.98} \\
\cmidrule(lr){1-7}
\end{tabular}
\end{table}

Compared to other methods, our proposed RoWSFormer demonstrates outstanding performance under rotation distortions. It achieves a PSNR exceeding 50 dB, indicating superior visual quality of the watermarked images even after rotation. Additionally, the extraction accuracy nearly reaches 100\% across all tested rotation angles, showcasing the model's exceptional robustness and effectiveness in handling rotational attacks. This significant improvement over existing methods highlights RoWSFormer's ability to maintain watermark integrity under challenging geometric transformations.

\subsubsection{Scaling} Scaling distortion involves scaling the image to different sizes. To improve the model’s resilience to changes in image size, we apply scaling factors randomly selected between 0.7 and 1.5 times the original dimensions during the training phase. In the testing phase, we evaluate the model’s robustness by using scaling factors ranging from 0.6 to 2. The experimental results for this scaling distortion are presented in \hyperref[table9]{Table IX}.

\begin{table}[ht]
\centering
\caption{PSNR and ACC with Different Ratio of Scaling}
\label{table9}
\begin{tabular}{>{\centering\arraybackslash}p{0.091\textwidth} >{\centering\arraybackslash}p{0.057\textwidth} >{\centering\arraybackslash}p{0.04\textwidth} >{\centering\arraybackslash}p{0.03\textwidth} >{\centering\arraybackslash}p{0.03\textwidth} >{\centering\arraybackslash}p{0.03\textwidth} >{\centering\arraybackslash}p{0.03\textwidth}}
\cmidrule(lr){1-7}
\multirow{2}{*}{Model} & \centering \multirow{2}{*}{PSNR(dB)} & \multicolumn{5}{c}{ACC(\%)} \\
\cmidrule(lr){3-7}
 & & $r$=0.5 & 0.7 & 1 & 1.5 & 2 \\
\cmidrule(lr){1-7}
HiDDeN \cite{Zhu2018} & 34.49 & 90.31 & 93.16 & 95.93 & 92.98 & 91.02  \\
TSDL \cite{Liu2019novel} & 35.13 & 75.03 & 82.30 & 85.96 & 82.13 & 75.90  \\
MBRS \cite{Jia2021mbrs} & 36.78 & 82.08 & 88.90 & 92.79 & 88.94 & 82.88  \\
WFormer \cite{Lou2024WFormer} & 42.35 & 90.23 & 95.92 & \textbf{100} & 96.11 & 90.53 \\
Ours & \textbf{47.56} & \textbf{97.06} & \textbf{99.57} & \textbf{100} & \textbf{100} & \textbf{99.85} \\
\cmidrule(lr){1-7}
\end{tabular}
\end{table}

The proposed RoWSFormer model ensures high-quality watermarked images, achieving a PSNR value exceeding 47 dB. In terms of robustness, the model consistently attains an extraction accuracy of over 97\%, demonstrating its practicality and reliability in handling scaling distortions in most scenarios. 

\subsubsection{Affine Attack} Affine attacks involve manipulating images through rotation, translation, scaling, and shearing transformations, which can significantly distort the embedded watermark and challenge its accurate extraction. During the training phase, we configure the affine parameters as follows: rotation angles randomly selected between \( -30^\circ \) and \( 30^\circ \), translations up to 0.1 in both horizontal and vertical directions, scaling factors set to 0.7 (reducing the image size by 30\%), and shearing angles between \( -30^\circ \) and \( 30^\circ \). 

In the testing phase, we assess the robustness by applying four distinct levels of affine attack strength, following the methodology outlined in WFormer \cite{Lou2024WFormer}. As illustrated in \hyperref[table10]{Table X}, we evaluated and compared our proposed RoWSFormer against other models under these varying affine attack intensities.

\begin{table}[ht]
\centering
\caption{PSNR and ACC with Different strengths of affine attacks}
\label{table10}
\begin{tabular}{>{\centering\arraybackslash}p{0.091\textwidth} >{\centering\arraybackslash}p{0.045\textwidth} >{\centering\arraybackslash}p{0.069\textwidth} >{\centering\arraybackslash}p{0.049\textwidth} >{\centering\arraybackslash}p{0.049\textwidth} >{\centering\arraybackslash}p{0.049\textwidth} }
\cmidrule(lr){1-6}
\multirow{3}{*}{Model} & \centering \multirow{3}{*}{PSNR(dB)} & \multicolumn{4}{c}{ACC(\%)} \\
\cmidrule(lr){3-6}
 & & $s$=(10, 0.1, 0.7, 30) & (0, 0.2, 0.7, 30) & (0, 0.1, 0.6, 30) & (0, 0.1, 0.7, 20)  \\
\cmidrule(lr){1-6}
HiDDeN \cite{Zhu2018} & 33.58 & 70.03 & 68.72 & 73.19 & 77.88 \\
TSDL \cite{Liu2019novel} & 33.14 & 68.16 & 65.57 & 70.11 & 68.96 \\
MBRS \cite{Jia2021mbrs} & 35.01 & 88.39 & 81.62 & 84.35 & 92.11 \\
WFormer \cite{Lou2024WFormer} & 36.91 & 91.79 & 89.63 & 93.32 & \textbf{100} \\
Ours & \textbf{44.16} & \textbf{99.08} & \textbf{99.98} & \textbf{99.15} & \textbf{100} \\
\cmidrule(lr){1-6}
\end{tabular}
\end{table}

The experimental results demonstrate that RoWSFormer significantly outperforms existing models in both imperceptibility and robustness metrics. Specifically, our model achieves a PSNR of 44.16 dB, which is over 7 dB higher than that of other methods, indicating superior visual quality of the watermarked images. Moreover, RoWSFormer maintains an extraction accuracy exceeding 99\%, showcasing its exceptional resilience to affine attacks. These results highlight the practical effectiveness of our model in real-world scenarios where such geometric distortions are common.

\subsection{Ablation Study}
\subsubsection{Effectiveness of LCEB} We remove the LCEB in LCESTB. With the help of LCEB, the PSNR and ACC of RoWSFormer improved by 1 dB and 0.5\% respectively. This may be because the LCEB introduces channel information and local information.

\subsubsection{Effectiveness of FEB} We remove the FEB in FETB. With the help of FEB, the PSNR and ACC of RoWSFormer improved by 0.5 dB and 0.3\% respectively. This may be because the FEB introduces frequency information.

\section{Conclusion}
In this paper, we introduce a robust image watermarking model named RoWSFormer. RoWSFormer leverages the Swin Transformer architecture offering enhanced flexibility in model design and effectively capturing multi-scale features of the image. Through extensive quantitative experiments, we demonstrate that RoWSFormer outperforms SOTA models in terms of invisibility and robustness, particularly against geometric attacks.

\bibliographystyle{IEEEtran}
\bibliography{IEEEabrv,RoWSFormer}

\end{document}